\documentclass[preprint,12pt]{elsarticle}

\usepackage{amssymb}
\usepackage{amsthm}
\usepackage{graphicx}
\usepackage{epsfig}
\usepackage{epstopdf}
\usepackage{bibentry}  
\usepackage{pdflscape}

\begin{document}
\journal{Physics Letters A}

\begin{frontmatter}
\title{On the Scale free laws of Urban Facilities}
\author{Liang Wu\footnote{Email: LiangWu@scu.edu.cn}, Yang Li and Xuezheng Chen} 



  

\begin{abstract}


We implement a double stochastic process as the mathematical model for the spatial point patterns of urban facilities. 
We find that the model with power covariance function can produce the best fit not only to $K$ function (whose derivative gives the radial distribution $\rho(t) = K'(t)/2\pi t$) but also to additional facts of spatial point patterns. These facts include the mean-variance relationship of number of events in a series of expanding bins, and other statistics beyond the first two orders, such as inter-event distribution function $H(t)$ and nearest neighbor distribution functions $G(t)$ and $F(t)$. 

\end{abstract}


\end{frontmatter}

\makeatletter

\section{Introduction}

The rapid urbanization becomes one of the predominant process in the human history. It arouses much interest to study cities and urban lives in the scientific communities\cite{Bettencourt,NatureUrbanGrowth}. Like many other physical systems, despite the complex underlying structure, macro statistical regularities such as power laws emerge\cite{Bettencourt,CityZipf}. With the increasing ability to collect the spatial coordinate data of facilities and buildings from electronic map providers, researchers begin to study the spatial substructure of cities. It is reported in \cite{liangwu} that $K$-function (whose derivative gives the radial distribution function $\rho(t) = K'(t)/2\pi t$) is a power function. Besides, the mean-variance relationship of number of events in a series of expanding bins is also a power function. These empirical results can be embedded in a double stochastic process model when the covariance function is a power function.  

Scientific interest in power law relations stems partly from the possibility that the power function might point to a deep origin in the dynamical process that generate the power law relation. 
Unfortunately, the detection and characterization of power laws are complicated by the large fluctuations that occur in the tail of the distribution and by the difficulty of identifying the range over which power law behavior holds\cite{clauset}. The fact that actual physical systems are finite also hinders the test of power laws which characterize long range interactions. As \cite{truthpower} points out that one should not only consider a detailed mechanism in driving dynamics, but also the extent of statistical support for a reported power law. The linear relationship on a log-log plot is clearly established for urban facilities in \cite{liangwu} over more than 2 orders of magnitude, however the considered statistics $K$-function and mean-variance relationship only capture the statistical properties of the first two orders. They do not give a complete picture. As an illustration of the insufficiency of second-order statistics, \cite{baddeley} describes a class of non-Poisson processes however for which $K(t)=\pi t^2$ coincides with the Poisson random process. 
Besides, the empirical evidence from $K$-function is not strong enough to support the power laws. Except for the power function, there are many other functions that can give good fit to $K(t)$. The power law rule would be on a more solid basis if the model estimated from the first and the second order statistics fits other statistics including higher order statistics.

Higher order statistics can be defined in terms of the joint intensity functions for the occurrence of specified configurations of three, four, etc. events. Interpretation would be difficult in practice since, for example, the third-order intensity functions of a stationary, isotropic process requires three arguments, the fourth-order function, five, and so on. There are several distribution functions which are easy to interpret. They can serve as additional statistics summary of spatial point process. These are $H(t)$ the distribution function of inter-event distance, $G(t)$ the distribution function of the distance from an arbitrary event to its nearest other event, and $F(t)$, the distribution function of the distance from an arbitrary point to its nearest event\cite{DiggleBook}. 

Except these statistics, mean-variance relationship can also serve as a statistical method to compare models. Denote $M$ and $V$ as the average and variance of number of events in a series of expanding bins, respectively. As shown in \cite{liangwu}, the power law rule of mean-variance relationship $V = aM^b$ is a natural result of a double stochastic model if the $K$-function is given as $K(t)=\pi t^2+K_0t^f$. Besides, their exponents are related by $b=1+f/2$.

The idea of this paper is to implement a double stochastic model which can generate random samples in resemblance to the actual point pattern of urban facilities. Then we propose some other covariance functions in addition to the power function to fit the $K$-function. We compare these models in their capability to fit additional second order statistics such as mean-variance relationship, and other statistics beyond the first two orders, such as inter-event distribution function $H(t)$ and nearest neighbor distribution functions $G(t)$ and $F(t)$.

\section{Choice of Covariance function}
A double stochastic process (DSP) model is introduced to model the spatial structure of urban facilities\cite{liangwu}. The DSP model assumes that there are two layers of stochastic process. The first layer stochastic process is a correlated random field of density function $\Lambda(x)$, which models the inhomogeneous concentration of facilities in a city. Conditional on the density, the location of urban facilities is based on a Poisson process. The DSP model simulates the growth process of urban facilities in the sense that urban facilities are evolving stochastically on top of an existing structure of a city, while the city structure itself follows a separate stochastic process.

A relatively flexible and tractable construction to encompass the non-negative constraint for density processes is log-Gaussian processes\cite{Moller}, i.e., the density function is drawn from a log-Gaussian random process $\Lambda(\vec{x}) = \exp(S(\vec{x}))$. $S$ is assumed to be a stationery isotropic Gaussian, $S(\vec{x}) \sim \mathcal{N}(\mu, \sigma^2)$. Its spatial dependence is given by its covariance density 
$\textnormal{Cov}(S(\vec{x}), S(\vec{x}+\vec{u})) = \sigma^2 r(|\vec{u}|)$. 
The first and second order statistics of $S$ and $\Lambda$ fields are related by
$m=\textnormal{E}[\Lambda(\vec{x})] = \exp(\mu+\sigma^2/2)$, and $\gamma(|\vec{u}|)=\textnormal{Cov}(\Lambda(\vec{x}), 
\Lambda(\vec{x}+\vec{u}))=\exp(2\mu + \sigma^2)[\exp(\sigma^2r(\vec{u}))-1]$. 

The radial distribution function $\rho(t)=(2\pi t)^{-1}K'(t)$ are related to the second order statistics of $S$ and $\Lambda(\vec{x})$ by,
\begin{eqnarray}
\sigma^2r(t) = \log(\rho(t)) \\
\gamma(t) = m^2(\rho(t)-1)  
\end{eqnarray} 

Now, we propose 3 models of the radial distribution function $\rho(t)$ as, 
\begin{eqnarray}
\label{eq:model1}
\rho_1(t) = 1+\theta_{11}(1+t^2)^{-\theta_{12}/2} \\
\label{eq:model2}
\rho_2(t) = 1+\exp{(\theta_{21}-\theta_{22}t)} \\
\label{eq:model3}
\rho_3(t) = \exp(\theta_{31}\exp{(-\theta_{32}t)})
\end{eqnarray}
These different models of $\rho(t)$ are hereafter referred as m1, m2 and m3 respectively. The radial distribution function $\rho_1(t)$ of m1 is actually a power function $\rho_1(t)\propto 1+\theta_{11}t^{-\theta{12}}$ at large values of $t$. 
1 is added to avoid the divergence at $t=0$. Note that for m1 $K(t) = \pi t^2 + \frac{\theta_{11}}{1-\theta{12}}t^{-\theta_{12}+1}$, which is the model suggested in \cite{liangwu}. Note that m2 is the first order expansion of m3 $\rho_3(t)\approx 1+\theta_{31}\exp(-\theta_{32}t)+O(\exp(-2\theta_{32}t))$. Therefore, the covariance of both m2 and m3 are exponential function, which dies quickly as $t$ increases and the interaction between events are short-ranged. 

\begin{figure}[htbp]
\begin{center}
\includegraphics[width=0.9\linewidth]{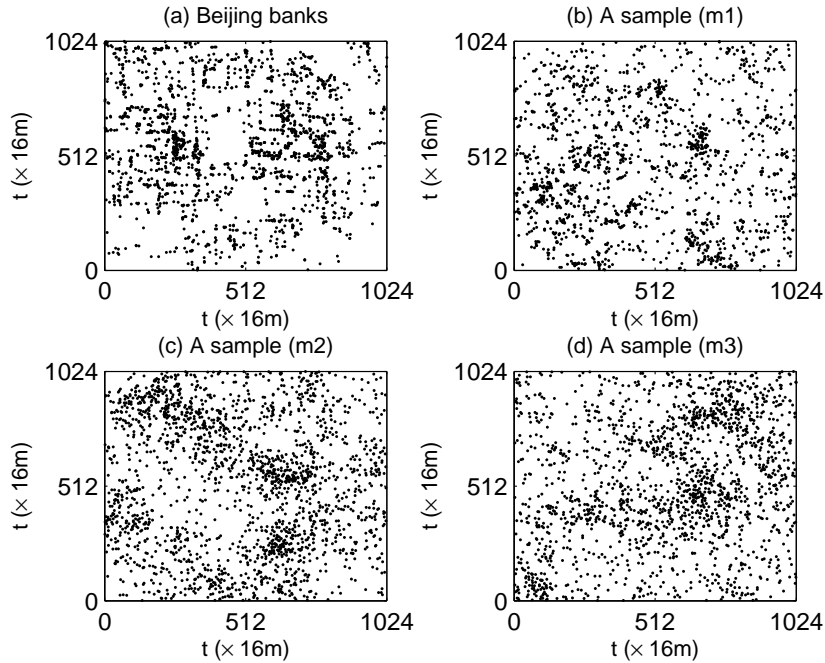}
\end{center}
\caption{\label{fig:scatter}
(a) Original locations of banks in a metropolitan area of Beijing ($2^{14} \times 2^{14} m^2$) mapped to a ($2^{10} \times 2^{10}$) lattice; A random sample of the point pattern generated from the DSP when the radial distribution function $\rho(t)$ is given by:  (b) Eq.~(\ref{eq:model1}); (c) Eq.~(\ref{eq:model2}); (d) Eq.~(\ref{eq:model3})
}
\end{figure}

\section{Results}
Similar to \cite{liangwu}, we take banks in Beijing as an illustrative example. We present the original spatial data of banks in Fig.~\ref{fig:scatter}(a). The data is constrained to the metropolitan area of Beijing, which covers $2^{14}\times 2^{14} m^2$. We map the data to a $2^{10}\times 2^{10}$ lattice by taking a transformation of the coordinate $x$ of each point as $x'=[x/2^4]$, rounded to the nearest integer. 

The estimation of $\rho(t)$ is not convenient in the spatial analysis of point patterns. Knowing that 
\begin{eqnarray}
K(t)=\int_0^t ds 2\pi s\rho(s) = m^{-1}\textnormal{E}[N_0(t)]
\end{eqnarray}
where $N_0(t)$ is the number of further events within distance $t$ of an arbitrary event, we estimate the models by fitting $K(t)$. 

\begin{figure}[htbp]
\begin{center}
\includegraphics[width=0.9\linewidth]{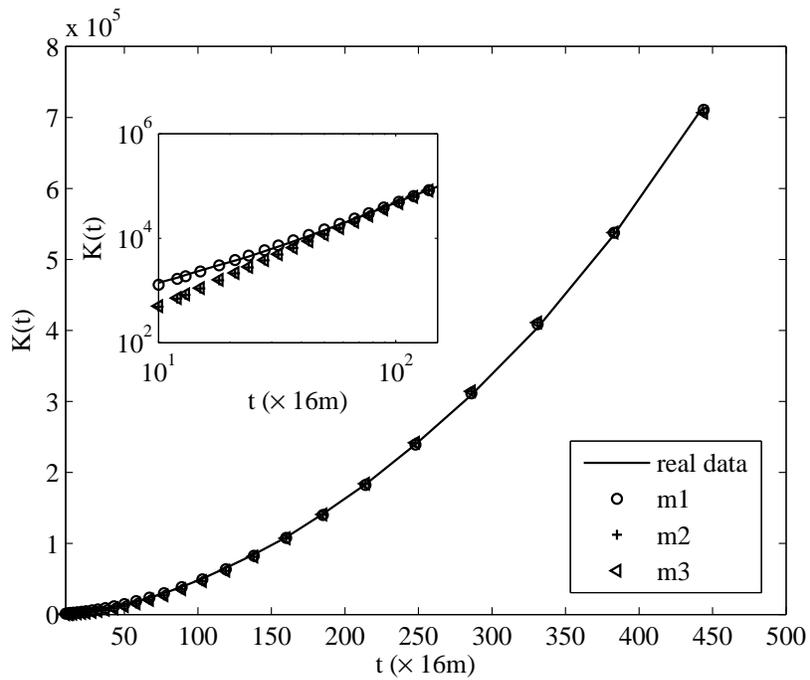}
\end{center}
\caption{\label{fig:1}
Fitting of $K(t)$ by 3 models. All three models fit the data quite well in the large range from 100 meters to 7000 meters. There is a slight difference when $r<300$ meters as one can see from the log-log plot in the sub-window. The log-log plot shows that m1 fits slightly better than the other two models. 
}
\end{figure}

Denote $\hat{K}(t)$ the estimator calculated from the data after edge correction is applied to the square window in our case, and $K(t,\theta)$ the theoretical K-function. $\theta$ is estimated to minimize the square error $J = \sum_{i=1}^I(\hat{K}(t_i)-K(t_i,\theta))^2/I$ between $\hat{K}(t)$ and $K(t,\theta)$. We choose 26 $t_i$s spanning from 5 to 500 units (each unit is 16 meters) which are separated in such a way that they are somehow uniformly distributed in log space to give more emphasis on small $t$s as shown in Fig~\ref{fig:1}. From the Figure, all three models fit the data quite well in the large range from 100 meters to 7000 meters. There is a slight difference when $t<300$ meters as one can see from the log-log plot in the sub-window. The log-log plot shows that m1 fits slightly better than the other two models. 

After we fit the models to $K(t)$, we can use the parameters to run simulations. Random samples can be generated with the Fourier filtering method\cite{LongRangeGenerator} when the radial distribution function (therefore the covariance function $\sigma^2r(t)$ of $S$ filed) is given. A random sample for each model of $\rho(t)$ is depicted in Fig.~\ref{fig:scatter}(b), Fig.~\ref{fig:scatter}(c) and Fig.\ref{fig:scatter}(d) respectively. By first checking the graphs with naked eyes, one may conclude that the point pattern generated by m1 matches closer to the actual pattern of banks than the other two models. 
Rigorous statistical checks are conducted in a later section of this paper.

In Fig.~\ref{fig:1}, we plot the fitting results of three models in one Figure. All three models are nearly indistinguishable in the large range of $t$ from 100 meters to 7000 meters. One can not really tell the difference in the regular plot. Although the covariance function for m2 and m3 are exponential which would deviate from the power function of m1 at large values of $t$, we can not reject them based on the fitting of $K(t)$ given the finite size of actual physical systems. Fortunately, there is slight difference when $r<300$ meters as one can see from the log-log plot in the sub-window of Fig.~\ref{fig:1}. The log-log plot shows that m1 fits slightly better than the other two models. 

\begin{figure}[htbp]
\begin{center}
\includegraphics[width=0.9\linewidth]{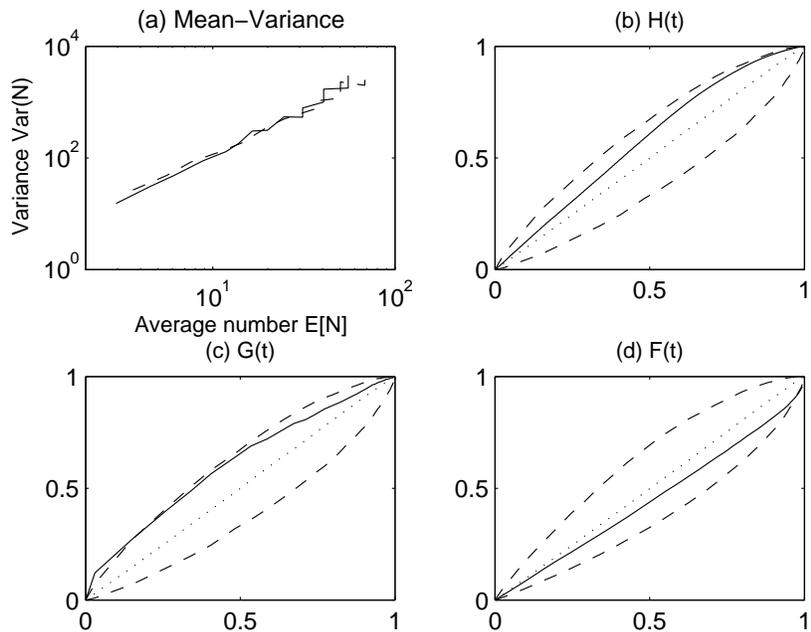}
\end{center}
\caption{\label{fig:21}
Statistical test of m1 using (a) mean-variance relationship; (b) $H(t)$ the distribution function of inter-event distance; (c) $G(t)$ the distribution function of the distance from an arbitrary event to the nearest other event; and (d) $F(t)$, the distribution function of the distance from an arbitrary point to the nearest event. In log-log plot (a) the solid line represents the average over 1,000 samples generated from model while the dashed line represents the real data. In QQ-plot (b), (c) and (d), the solid lines represent the empirical distribution function(EDF) of the real data against the theoretic distribution function(TDF) estimated from simulation samples; dashed lines represent the 1st and 99th centiles of the simulated data; dotted lines represent the median.
}
\end{figure}

\begin{figure}[htbp]
\begin{center}
\includegraphics[width=0.9\linewidth]{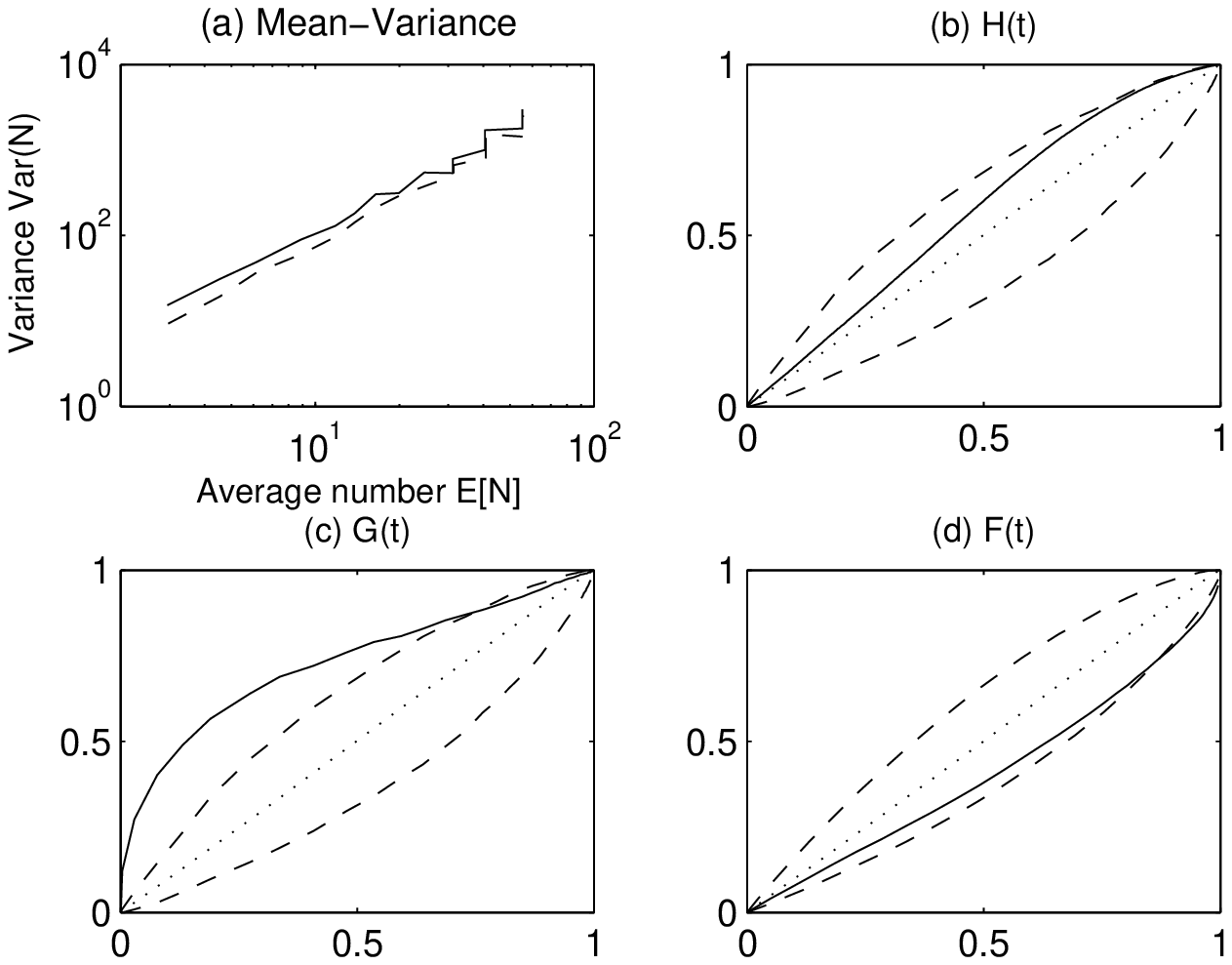}
\end{center}
\caption{\label{fig:22}
Statistical test of m2 using (a) mean-variance relationship; (b) $H(t)$ the distribution function of inter-event distance; (c) $G(t)$ the distribution function of the distance from an arbitrary event to the nearest other event; and (d) $F(t)$, the distribution function of the distance from an arbitrary point to the nearest event. In log-log plot (a) the solid line represents the average over 1,000 samples generated from model while the dashed line represents the real data. In QQ-plot (b), (c) and (d), the solid lines represent the empirical distribution function(EDF) of the real data against the theoretic distribution function(TDF) estimated from simulation samples; dashed lines represent the 1st and 99th centiles of the simulated data; dotted lines represent the median.
}
\end{figure}

\begin{figure}[htbp]
\begin{center}
\includegraphics[width=0.9\linewidth]{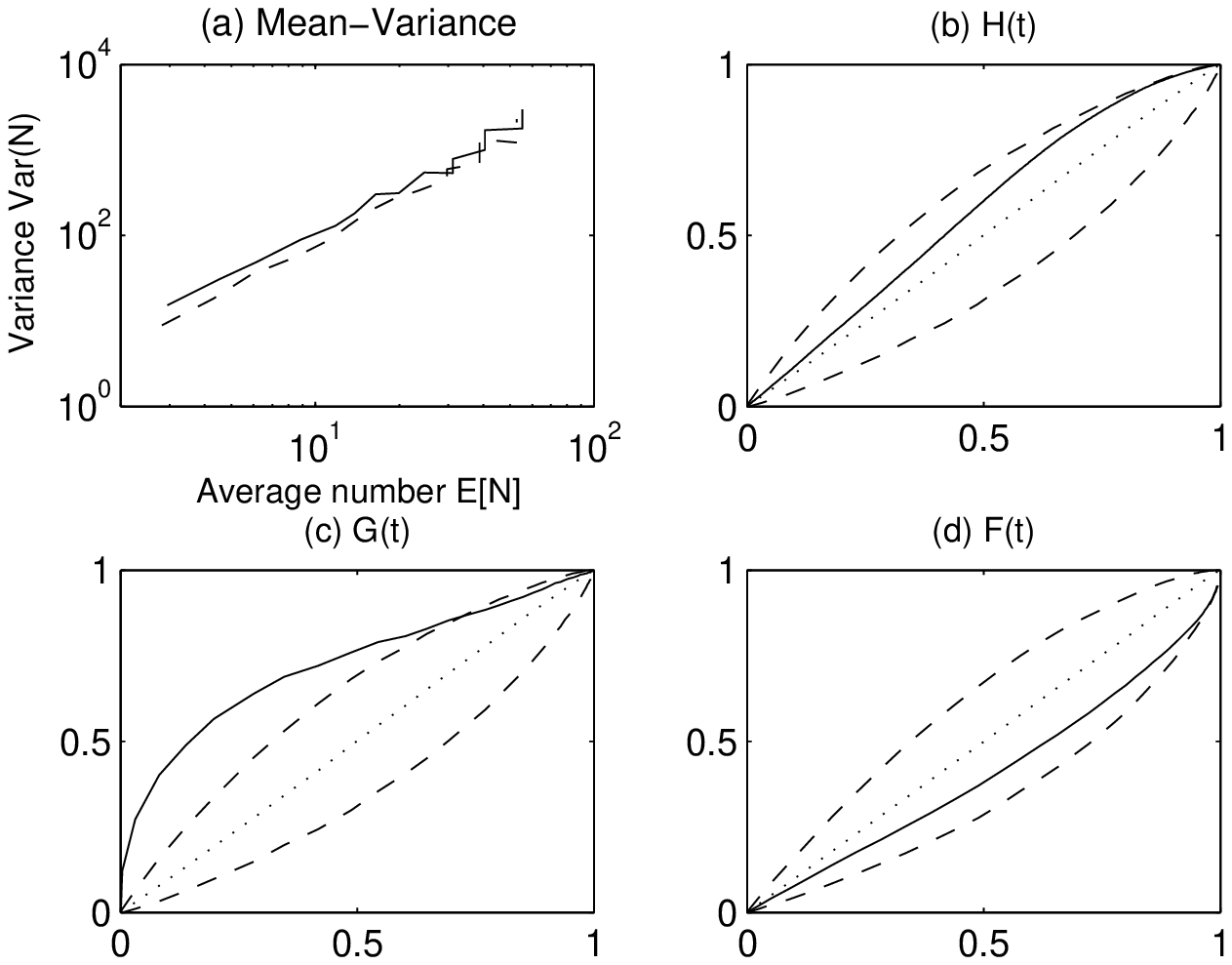}
\end{center}
\caption{\label{fig:23}
Statistical test of m3 using (a) mean-variance relationship; (b) $H(t)$ the distribution function of inter-event distance; (c) $G(t)$ the distribution function of the distance from an arbitrary event to the nearest other event; and (d) $F(t)$, the distribution function of the distance from an arbitrary point to the nearest event. In log-log plot (a) the solid line represents the average over 1,000 samples generated from model while the dashed line represents the real data. In QQ-plot (b), (c) and (d), the solid lines represent the empirical distribution function(EDF) of the real data against the theoretic distribution function(TDF) estimated from simulation samples; dashed lines represent the 1st and 99th centiles of the simulated data; dotted lines represent the median.
}
\end{figure}

We extend our analysis to include additional statistics to further test the goodness of fit of different models. The first one is the mean-variance relationship as reported in \cite{liangwu} which is still the second order statistics. Denote $M$ and $V$ as the average and variance of number of events in a series of expanding bins. They are shown in \cite{liangwu} that the power law $V = aM^b$ is a natural result if $\rho(t)=1+\theta_1 t^{\theta_2}$. We plot the mean-variance relationship in Fig.~\ref{fig:21}(a), Fig.~\ref{fig:22}(a), and Fig.~\ref{fig:23}(a) for m1, m2 and m3 respectively. The solid line in each figure is obtained by taking average over 1,000 samples. The dashed line is obtained from the real data. We can see that m1 generates the best fit.

The other three statistical tests are based on: $H(t)$ the distribution function of inter-event distance, $G(t)$ the distribution function of the distance from an arbitrary event to the nearest other event, and $F(t)$ the distribution function of the distance from an arbitrary point to the nearest event\cite{DiggleBook}. Take $H(t)$ as an example, conventional approach to assess the model is to use QQ-plot to compare the empirical distribution functions(EDF) $\hat{H}_0(t)$ of data with the theoretical distribution function(TDF). In our case, it is not straightforward to compute the TDF given the complexity of double stochastic models. We therefore proceed to estimate the TDF from samples. We calculate EDF's $\hat{H}_i(t),i=1,2,...,s=1,000$ from each of $s$ independent simulations. 
An estimation of the TDF $\hat{H}(t) = \sum_{i=1}^s\hat{H}_i(t)/s$. 
With 1,000 samples, we can also estimate the centiles and median of the sampled EDFs. If the real data can be explained by the model, it can be thought of as a random sample and therefore the EDF $\hat{H}_0(t)$ of the real data should be enclosed in the envelope formed by the 1st and 99th centiles of the sampled EDFs with probability 98\%.

We plot the EDF $\hat{H}_0(t)$ against the TDF $\hat{H}(t)$ with solid lines in Fig.~\ref{fig:21}(b), Fig.~\ref{fig:22}(b), and Fig.~\ref{fig:23}(b) for m1, m2 and m3 respectively. In addition we plot the medians of the sampled EDFs against the TDF with dotted lines. Note that the TDF is the mean of the sample EDFs. The median of the sampled TDF is almost a straight line against the TDF in the Figure, which suggests that the distribution of the sampled EDFs could be somehow symmetric. The results for $G(t)$ and $F(t)$ are prepared similarly and plotted in sub-figure (c) and (d). 


We can see from each subfigure (b), $\hat{H}_0(t)$ is enclosed in 1st and 99th envelopes of the simulation data, which means that all three models pass the test. The distribution function $H(t)$ of inter-event distance captures the overall characteristics of events over relative long distance. They produce similar results. However, both m2 and m3 fail the test if $G(t)$ is involved. The actual data has more nearest neighbors than m2 and m3 predict at a given distance away from each event since $\hat{G}_0(t)$ is much larger than the 99th centile of the sampled EDF when $\hat{G}_0$ is less than 0.5. The problem is not so severe for $F(t)$. However, $\hat{F}_0(t)$ lies close to, albeit above the 1st centile of the sample EDF generated by m2. The events generated from m2 are more closer to random than the other 2 models due to the fact that the covariance function of m2 decreases most quickly among 3 models. Model 1 passes the test for $F(t)$. Although it barely passes for $G(t)$, it is still the best among 3 models. The actual data is more strongly clustered than the power law predicts at small distance. Small correction needs to be made to explain the small distance behaviors. 

\begin{figure}[htbp]
\begin{center}
\includegraphics[width=0.9\linewidth]{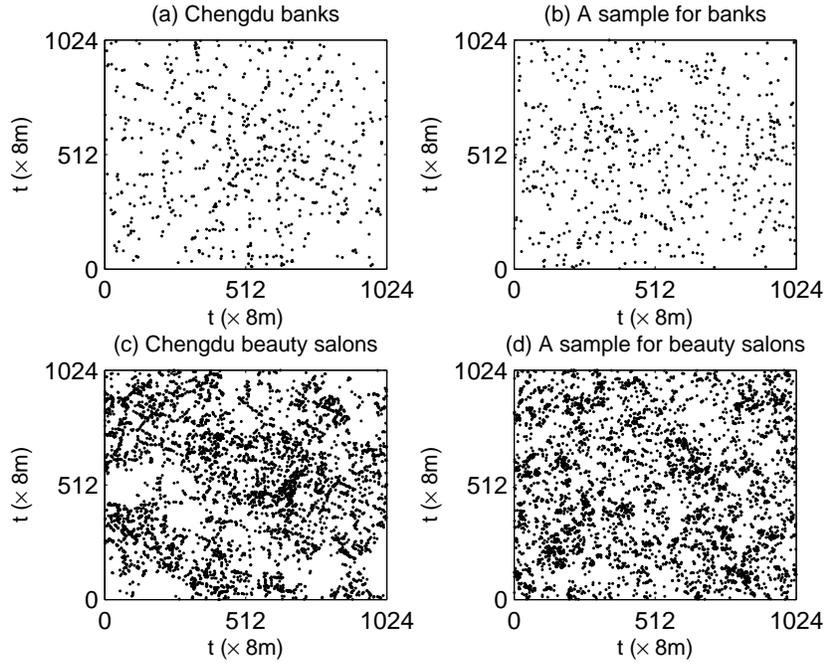}
\end{center}
\caption{\label{fig:chengdu}
(a) and (c) depict the original locations of banks and beauty salons in a metropolitan area of Chengdu ($2^{13} \times 2^{13} m^2$) mapped to a ($2^{10} \times 2^{10}$) lattice, respectively; (b) and (d) depict a random sample of the corresponding point pattern of banks and beauty salon in Chengdu, respectively. The sample is generated from the DSP when the radial distribution function $\rho(t)$ is estimated as the power function Eq.~(\ref{eq:model1}).
}
\end{figure}

We have extended out tests to include extra facilities such as pharmacy, convenient stores, and beauty salons and also to another big city Chengdu located in Southwestern part of China. The results are similar. The DSP model whose covariance function is a power function always produces the best fit. We present the results of two facilities in Chengdu as representative examples. The original data appears to have vast distinct morphology mainly due to their different incidences in the metropolitan area of Chengdu ($2^{13} \times 2^{13} m^2$ rectangular region around the center). The original point pattern and a sample is plotted in Fig.~\ref{fig:chengdu} for Chengdu banks and Chengdu beauty salons, respectively. As one can see from the Figure, the DSP model with power covariance function can generate point patterns in good resemblance with the original data. Once might notice the latent circular pattern in the real data of Chengdu (compared to the rectangular pattern in that of Beijing), which is due to the circular city planning and road construction. It is an unnecessary detail which is not captured with the current homogeneous model. The statistical comparison is not presented since it involves too many figures, but can be given upon request. 

\section{Conclusion and Discussion}
It is interesting to observe the power law relations in a system because they might point to a deep origin in the dynamical process that generates the power law relation. 
Unfortunately, the detection and characterization of power laws are complicated by the large fluctuations that occur in the tail of the distribution and by the difficulty of identifying the range over which power law behavior holds. The fact that actual physical systems are finite also hinders the test of power laws which characterize long range interactions.

The linear relationship on a log-log plot is clearly established for urban facilities in \cite{liangwu} over more than 2 orders of magnitude. In this paper, we implement a method to test the validity of power laws by showing that a double stochastic model whose power law covariance function is estimated from the first two orders of statistics can give the best fit to additional statistics. The statistics includes mean-variance relationship, and other statistics beyond the first two orders, such as inter-event distribution function $H(t)$ and nearest neighbor distribution functions $G(t)$ and $F(t)$. 

It should be noted that we assume that the density field $S(\vec{x})$ of the double stochastic model is stationery and isotropic. This condition is violated in practice in that first the city is often developed around a center and the concentration of facilities is expected to gradually drop from the center to the outlying areas of the city, and secondly the occupation of parks and other big buildings like the Forbidden City in Beijing create a vacuum where no facilities can be found. Despite the inadequacy, the model with only 2 parameters can still give good fit to many statistics, some of which are beyond the first two orders.

\section{Acknowledgments}

We gratefully acknowledge the following financial supports: the Start-up Funds from Sichuan University under Grants No. yj201322 and No. yj201353; the Fundamental Research Funds for the Central Universities under Grants No. skyb201403, No. skyb201404 and No. Skzx2015-sb58.

\end{document}